\documentclass[aps,prb,twocolumn,showpacs,superscriptaddress,amsmath,amssymb]{revtex4-1}
\usepackage{graphicx}  
\begin{document}
\title{Long-range interactions between substitutional nitrogen dopants in graphene: electronic properties calculations}
\author{Ph. Lambin}
\affiliation{Physics Department (PMR), University of Namur (FUNDP), B-5000 Namur, Belgium}
\author{H. Amara}
\affiliation{Laboratoire d'Etude des Microstructures, ONERA-CNRS, BP 72, 92322 Ch\^atillon Cedex, France}
\author{F. Ducastelle}
\affiliation{Laboratoire d'Etude des Microstructures, ONERA-CNRS, BP 72, 92322 Ch\^atillon Cedex, France}
\author{L. Henrard}
\affiliation{Physics Department (PMR), University of Namur (FUNDP), B-5000 Namur, Belgium}

\date{\today}

\begin{abstract}
Being a true two-dimensional crystal, graphene has special properties. In
particular, a point-like defect in graphene may induce perturbations in the long
range. This characteristic questions the validity of using a supercell geometry
in an attempt to explore the properties of an isolated defect. Still, this
approach is often used in {\it ab-initio} electronic structure calculations, for
instance. How does this approach converge with the size of the supercell is
generally not tackled for the obvious reason of keeping the computational load
to an affordable level.  The present paper addresses the problem of
substitutional nitrogen doping of graphene. DFT calculations have been performed
for $9\times 9$ and $10\times 10$ supercells. Although these calculations
correspond to N concentrations that differ by $\sim10$\%, the local densities of
states on and around the defects are found to depend significantly on the
supercell size. Fitting the DFT results by a tight-binding Hamiltonian makes it
possible to explore the effects of a random distribution of the substitutional N
atoms, in the case of finite concentrations, and to approach the case of an
isolated impurity when the concentration vanishes. The tight-binding Hamiltonian
is used to calculate the STM image of graphene around an isolated N atom. STM
images are also calculated for graphene doped with 0.5~at\% concentration of
nitrogen. The results are discussed in the light of recent experimental data and
the conclusions of the calculations are extended to other point defects in
graphene.
\end{abstract}
\maketitle

\section{Introduction}
Local defects and chemical doping is a well-documented way to tune the
electronic properties of graphene.~\cite{Niyogi2011} More specifically, the
benefits of doping have been underlined in the context of
(bio)sensing,~\cite{Wang2010} lithium incorporation
battery,~\cite{Reddy2010,Li2012} and in other fields. Nitrogen (N) is a natural
substitute for carbon in the honeycomb structure due to both its ability to form
sp$^2$ bonds and its pentavalent character. It is not a surprise, then, that
many publications deal with the production of N-doped graphene realized either
by direct growth of modified layers~\cite{Zhao2011,Wei2009} or by post-synthesis
treatments.~\cite{Guo2010,Joucken2012} Doping single-wall carbon nanotubes has
also been considered.~\cite{Arenal2010,Ayala2010,Lin2011}

Recent STM-STS experiments of N-doped graphene~\cite{Zhao2011,Joucken2012} have
demonstrated the occurrence of chemical substitution. These experiments provide
us with a detailed and local analysis with sub-atomic resolution of the
electronic properties of the doped material. The STM images display a pattern
having three-fold symmetry around the N atoms and having a strong signal on the
C atoms bonded to the dopant, which {\it ab-initio} simulations reproduce
well.~\cite{Zheng2010} STS measurements have revealed a broad resonant
electronic state around the dopant position and located at an energy of 0.5~eV
above the Fermi level.~\cite{Joucken2012}

It is important to understand the effects of local defects and chemical doping
on the global electronic properties of graphene, on its quantum transport
properties and on its local chemical reactivity. It is the reason why the
electronic properties of graphene doped with nitrogen have been investigated by
several groups, mainly on the basis of {\it ab-initio} DFT
techniques.~\cite{Latil2004,Adessi2006,Lherbier2008,Zheng2010,Zhao2011,Fujimoto2011,Brito2012}
The central advantage of the {\it ab-initio} approach is to be parameter free. A
disadvantage is to be restricted to periodic systems as long as fast Fourier
transforms need to be used to link direct and reciprocal spaces. In most
instances, doping is therefore addressed in a supercell geometry that precludes
the study of low doping concentration or disorder.  In the case of single-wall
carbon nanotubes, however, the electronic properties of {\em isolated defects}
have been studied by first-principle methods based on scattering
theory.~\cite{Choi2000} For a nitrogen impurity in the (10,10) armchair
nanotube, the local density of states on the N atom shows a broad peak centered
at 0.53~eV above the Fermi level. The charge density associated with that
quasi-bound state (donor level in the language of semiconductor physics) extends
up to $\sim$1~nm away from the defect. This means that N atoms located
$\sim$2~nm apart can interact, which requires examining with care the intrinsic
validity of a supercell method.

The present work, based on both {\it ab-initio} DFT and semi-empirical
tight-binding electronic structure calculations, aims at looking for
interference effects generated by a distribution of N dopants in graphene as
compared to the case of an individual impurity. We resort to different
tight-binding parametrization strategies. The simplest one, based on just one
adjustable parameter related to the defect, permits analytical calculations of
the impurity levels. This model is described in Appendix A. We favor a more
realistic approach, in which the perturbation induced by the defect is allowed
to extend on the neighboring sites. This latter model is used to study the
effects of disorder on the local and global electronic properties of doped
graphene and to calculate STM images that we compare with available experimental
data.  The calculations and discussions are developed here for substitutional
nitrogen. The conclusions would be qualitatively the same for boron doping and
for other types of point-like defects.

\section{Methodology}
The SIESTA package~\cite{SanchezPortal1997} was used for the {\it ab-initio} DFT
calculations. The description of the valence electrons was based on localized
pseudo-atomic orbitals with a double-$\zeta$ polarization
(DZP).~\cite{Artacho1999} Exchange-correlation effects were handled within local
density approximation (LDA) as proposed by Perdew and Zunger.~\cite{Perdew1981}
Core electrons were replaced by nonlocal norm-conserving
pseudopotentials.~\cite{Troullier1991} Following previous
studies,~\cite{Zheng2010,Joucken2012} the first Brillouin zone (BZ) was sampled
with a $15\times 15$ grid generated according to the Monkhorst-Pack
scheme~\cite{Monkhorst1976} in order to ensure a good convergence of the
self-consistent electronic density calculations. Real-space integration was
performed on a regular grid corresponding to a plane-wave cutoff around
300~Ry. All the atomic structures of self-supported doped graphene have been
relaxed. 

As mentioned above, the DFT calculations are based on a code that requires a
periodic system. As a consequence, a supercell scheme was adopted to handle
substitutionaly doped graphene. The atomic concentration of dopants is then
directly related to the size of the supercell: 0.6~at\% for a $9\times 9$
supercell and 0.5~at\% for a $10\times 10$ supercell.

For the tight-binding parametrization, the following, well-established procedure
was followed.~\cite{Latil2004,Adessi2006,Lherbier2008} The Hamiltonian included
the $\pi$ electrons only, the hopping parameter between nearest-neighbor atoms
was set to the {\it ab-initio} DFT value $\gamma_0$ = -2.72~eV. The same hopping
was used between the N atom and its C first neighbors, which is not a severe
approximation. It is indeed shown in Appendix A that a non-diagonal perturbation
of the Hamiltonian can be offset by a renormalization of the N on-site energy,
and this energy is a free parameter to be adjusted to the results of the {\it
  ab-initio} calculations. The C on-site energies were assumed to vary with the
distance $d$ to the impurity according to a Gaussian law:
\begin{equation}
\varepsilon(d) = \varepsilon_C - |U|\,\exp(-0.5 d^2/\sigma^2) 
\label{gaussian}
\end{equation}
where $|U|$ is the depth of the potential well induced by the nitrogen (boron
induces a potential hump, instead~\cite{Latil2004,Khalfoun2010} ) and
$\varepsilon_C$ is the asymptotic bulk value of the on-site parameter of
carbon. $\varepsilon_C$ defines the Dirac energy of pure graphene and has no
other influence on the density of states (DOS) than a systematic shift of all
the electronic levels. The standard deviation $\sigma$ in eq.~(\ref{gaussian})
was set to 0.15~nm, according to data of Ref.~[\onlinecite{Khalfoun2010}],
whereas $U$ was chosen so as to best fit DFT local DOS calculated on the N atom
in the $9\times 9$ supercell (see below).

In order to avoid the constraints of a periodic tight-binding Hamiltonian, the
local DOS have been computed with the recursion method.~\cite{Haydock1975} 150
pairs of recursion coefficients were calculated and extrapolated to their
asymptotic values related to the edges $\varepsilon_C \pm 3\gamma_0$ of the
$\pi$ bands of pure graphene. It is worth emphasizing that this procedure avoids
to introduce any broadening of the energy levels. A drawback is the presence of
wiggles that may be observed in some densities of states. They are sorts of
Gibbs oscillations generated by the Van Hove singularities, in particular by the
abrupt discontinuities of the graphene $\pi$ DOS at both band edges.

\section{Periodic distribution of N} \label{section_periodic}
\begin{figure}
\centerline{\includegraphics*[width=8.5cm]{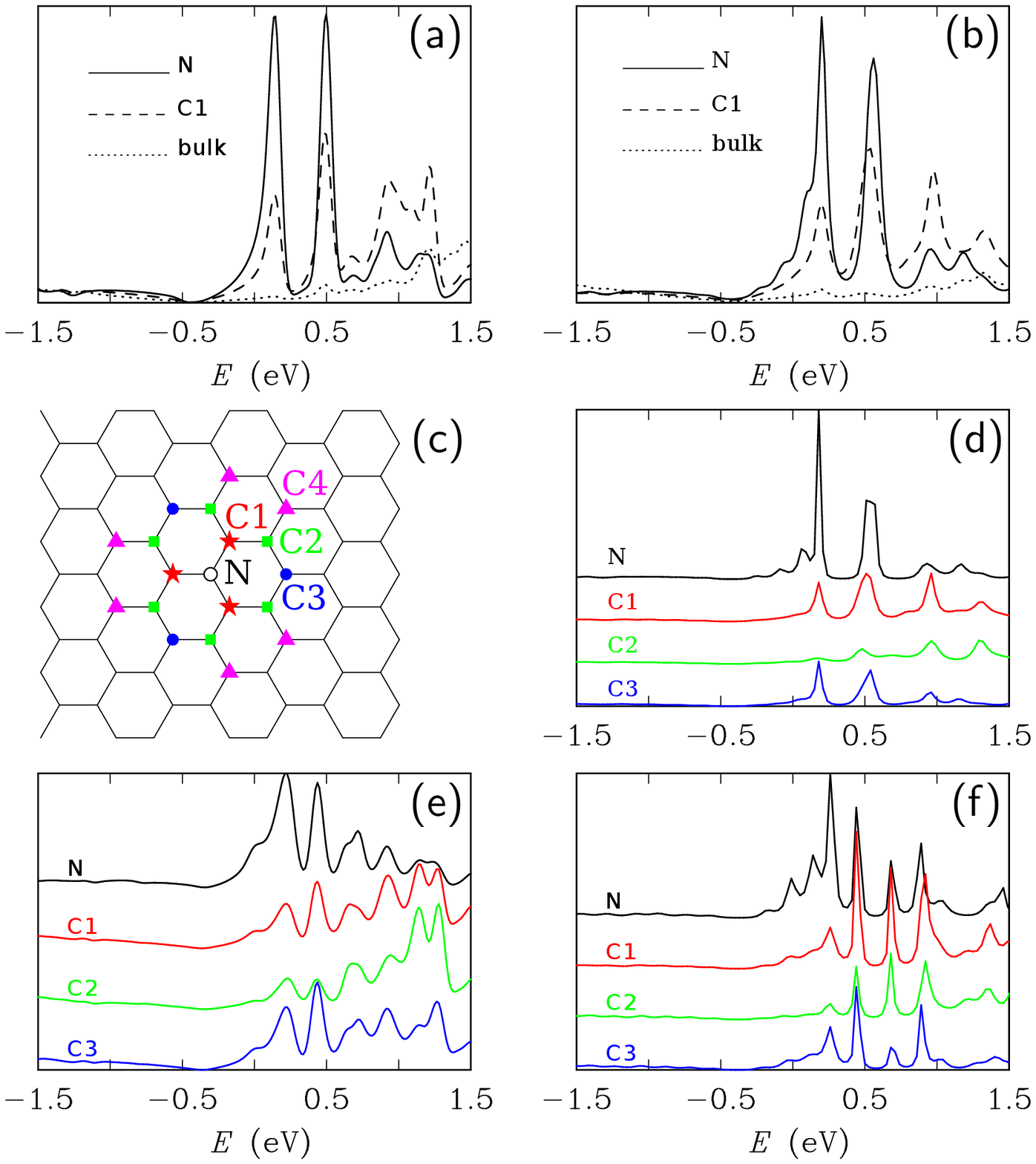}}
\caption{(a) DFT and (b) $\pi$ tight-binding local densities of states (LDOS) of
  graphene doped with nitrogen at a concentration of one impurity per $9\times
  9$ supercell (162 atoms). The different curves represent the local DOS on the
  N atom, on the first-neighbor carbons (C1), and on a few distant carbons
  (bulk). (c) Labels assigned to the nitrogen dopant (N) and its first, second,
  third, and fourth neighbors (C1, C2, C3, and C4). (d) Tight-binding LDOS of
  the $9\times 9$ superstructure (same as in panel b) without energy
  broadening. (e) DFT and (f) tight-binding LDOS on and around a substitutional
  N atom in graphene with $10\times 10$ periodic distribution (200 atoms per
  cell). The curves, shifted vertically for clarity, correspond to the atomic
  sites N, C1, C2, and C3 shown in panel (c). For all LDOS curve plotted in this
  figure, the zero energy corresponds to the Fermi level.}
\label{scdos}
\end{figure}

Fig.~\ref{scdos}(a) shows DFT local densities of states on the N atom, on the
first-neighbor atoms (C1) and on carbons located farther (0.5 -- 0.7~nm) away
(bulk), in a $9\times 9$ supercell. A broadening of 0.05~eV on a $45\times 45$
grid was used for the calculations.  The important observation is the existence
of two peaks, located at 0.55 and 0.92~eV above the minimum of the density of
states in the N local DOS.~\cite{Joucken2012}

The present DFT calculations are in overall good agreement with results
published for $4\times 4$,~\cite{Latil2004,Fujimoto2011} $5\times
5$,~\cite{Brito2012} $7\times 7$,~\cite{Zhao2011} and $10\times 10$
supercells.~\cite{Zheng2010} The double-peak structure in the unoccupied part of
the local N DOS plotted in Fig.~\ref{scdos}(a)) was not observed in previous
calculations due to the large energy broadening that was used
there,~\cite{Zheng2010,Zhao2011} except in Ref.~[\onlinecite{Joucken2012}]. It
will be demonstrated below that the observed double-peak structure is a
consequence of long-range interactions between the N impurities that were
reproduced periodically in the graphene sheet because of the supercell geometry.

Fig.~\ref{scdos}(b) shows tight-binding densities of states of the same $9\times
9$ supercell after Lorentzian broadening aimed at facilitating the comparison
with the DFT results. The latter are qualitatively well reproduced by the
$\pi$-electron tight-binding Hamiltonian by taking $|U|$ = 4~eV in
eq.~(\ref{gaussian}). The present work aims at tailoring a simple tool for
exploring the effects of the defect configurations. For that reason, achieving
the best possible fit of DFT calculations for a specific supercell is
superfluous. By comparison, the on-site energies in
Ref.~[\onlinecite{Adessi2006}] can be approximated by a sum of two Gaussians,
$\varepsilon(d) = \varepsilon_C - 2.95\,\exp(-0.5 d^2/\sigma_1^2) -
0.59\,\exp(-0.5 d^2/\sigma_2^2)$ (in eV), with $\sigma_1$ = 0.10~nm and
$\sigma_2$ = 0.39~nm. The second term has a longer range than the single
Gaussian law used in the present calculations, and the potential well at the N
location is about 10\% shallower than here.

The tight-binding densities of states of the same $9\times 9$ superstructure,
now calculated without energy broadening, are displayed in
Fig.~\ref{scdos}(d). A substructure clearly emerges where the broadened local
DOS only showed the double-peak feature discussed here above. In addition,
Fig.~\ref{scdos}(d) reveals that there are much less electronic states in the
region between 0 and 1.5~eV on the second-neighbor atoms (C2) than on the first-
(C1) and third-neighbor (C3) carbons (see Fig~\ref{scdos}(c) for the notations).
The same observation as for C2 can be made for the local DOS on the fourth
neighbors (not shown). This alternation between lower and higher densities of
states brings out the fact that the two sublattices of graphene are differently
affected by a substitutional defect (see Appendix A).

The reliability of the tight-binding calculations compared to DFT can be
appreciated from Figs.~\ref{scdos}(e,f) obtained for a $10\times 10$ supercell
(0.5~at\% concentration of N). The good agreement between the two approaches
reinforces the fact that the parametrization of the tight-binding on-site
energies used for the $9\times 9$ supercell describes well the doped graphene
systems and does not require further adjustment depending on the actual
concentration. Electronic states localized on and near the N dopant are observed
within 1.5~eV from the Dirac energy, as for the $9\times 9$ supercell. However,
the multi-peak structure in the local N DOS of $10\times 10$
(Fig.~\ref{scdos}(f)) is totally different from that of $9\times 9$
(Fig.~\ref{scdos}(d)), despite similar N concentrations (0.5~at\% and 0.6~at\%,
respectively). This finding is a clear indication of the importance of
interference effects among the supercells.

\begin{figure}
\centerline{\includegraphics*[width=6cm]{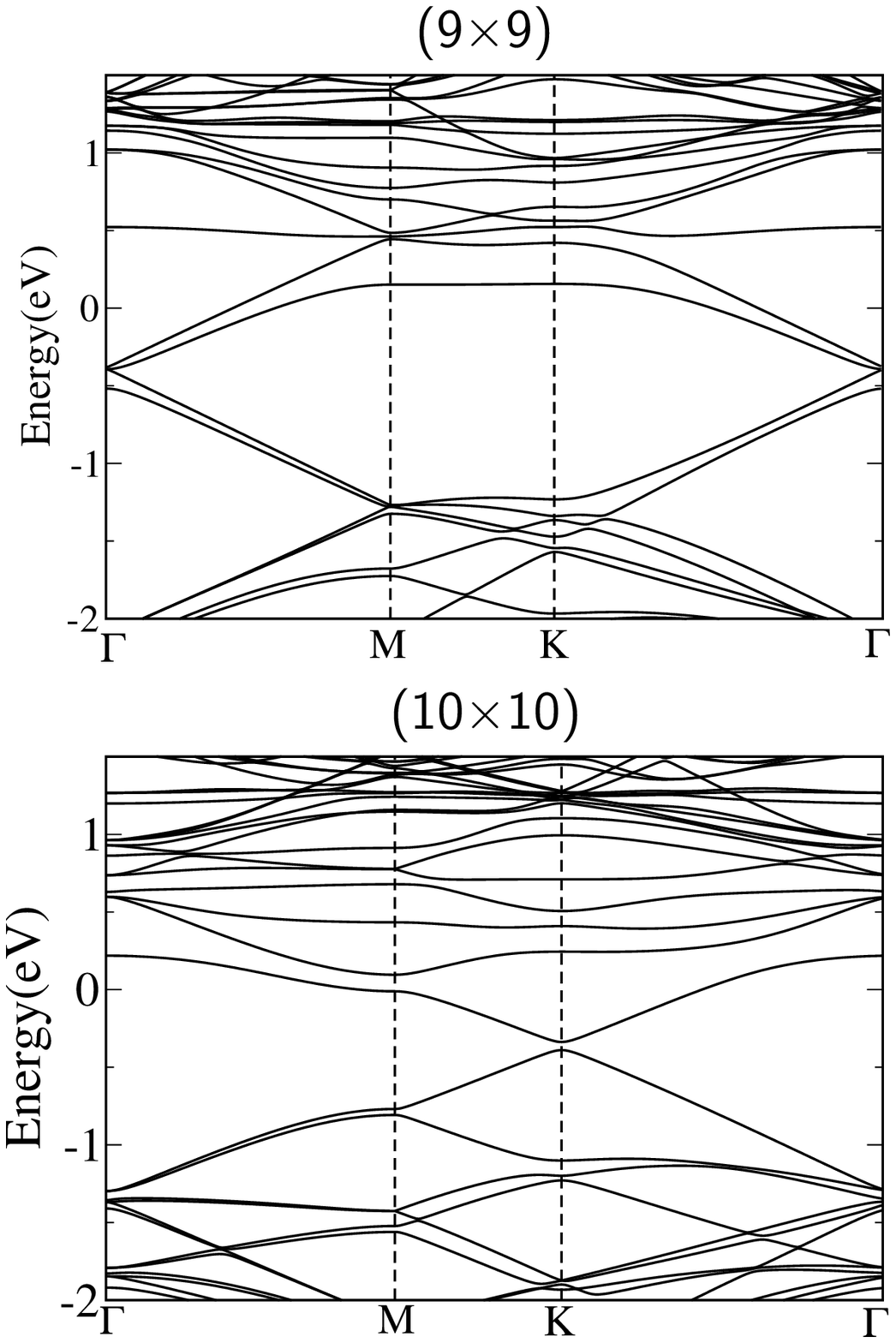}}
\caption{DFT band structures of the $9\times 9$ (top) and $10\times 10$
  (bottom) N-doped superstructures shown along the high-symmetry lines of their
  first Brillouin zone. The zero of energy coincides with the Fermi level.} 
\label{bstrdft}
\end{figure}

The band structures of the DFT calculations shown in Fig.~\ref{bstrdft} provide
us with another difference between the two superstructures: the $10\times 10$
N-doped superstructure has a direct band gap of $\sim$0.05~eV at the $K$ point
of its Brillouin zone, whereas the $9\times 9$ N-doped superstructure has no
gap. Symmetry considerations and arguments from perturbation theory developed in
Appendix B explain why it is so. The $K$ and $K'$ points of graphene move to the
$K$ and $K'$ points of the folded Brillouin zone of a $p\times p$ supercell when
$p$ is not divisible by 3, whereas they are both mapped onto the $\Gamma$ point
when $p$ is an integer multiple of 3.~\cite{Martinazz2010} In the latter case,
the fourfold degeneracy of the Dirac energy at the $\Gamma$ point of a supercell
of pure graphene is partly lifted by the perturbation brought about by the N
atoms. As demonstrated in Appendix B, there remain $\pi$ and $\pi^*$ bands that
cross the Dirac energy at $\Gamma$. This crossing as clearly visible in
Fig.~\ref{bstrdft}(a) for the $9\times 9$ superstructure, which therefore has no
gap. In direct space, a doped superstructure has point group symmetry
$D_{3h}$. The symmetry of the $\Gamma$ point of the superstructure, $D_{3h}$,
which is the same as that of the $K$ and $K'$ points of pure
graphene,~\cite{Malard2009,Kogan2012} has irreducible representations of
dimension one and two. A twofold degeneracy of some electron energy bands is
therefore allowed at the zone center of a doped superstructure. By contrast,
this degeneracy is forbidden at the $K$ and $K'$ points of the folded zone
because the symmetry of theses points, from $D_{3h}$ in pure graphene, is
reduced to $C_{3h}$ in the doped superstructure. The latter point group has only
one-dimensional irreducible representations. As a consequence, the crossing of
$\pi$ and $\pi^*$ bands is symmetry forbidden at $K$ and $K'$. A periodic
substitution of N for C therefore opens a gap in these $p\times p$ supercells
for which $p$ is not divisible by 3, as shown in Fig.~\ref{bstrdft}(b) for the
$10\times 10$ superstructure. It is demonstrated in Appendix B that the band gap
in those superstructures scales with $p$ like $E_g \approx V/p^2$, where $V$ is
of the order of 10~eV. Other local gaps of the order of $V/p^2$ appear in the
band structures of Figs.~\ref{bstrdft}(a,b) at the edges of the folded Brillouin
zone. They produce pseudo-gaps and sub-peaks in the densities of states clearly
discernible in Fig.~\ref{scdos} for both the $9\times 9$ and $10\times 10$
superstructures. As a result, supercells of the order of $30\times 30$ at least
would be necessary to reproduce the characteristic features of an isolated
impurity with an energy resolution better than $0.01$~eV.

In the tight-binding calculations, the band gap of the $10\times 10$
superstructure is located 0.14~eV below the Dirac point of pure graphene
($\varepsilon_C$). The Fermi level $E_F$ of the doped superstructures was found
to lie 0.27~eV and 0.25~eV above $\varepsilon_C$ for the $9\times 9$ and $10
\times 10$ systems, respectively, which means 0.43~eV and 0.39~eV above the
minimum of the DOS (the corresponding DFT values are 0.42 and 0.36~eV,
respectively). The $\sigma$ bands below the Fermi energy are completely filled,
the $\pi$ bands of the supercell must accommodate an extra electron brought
about by the nitrogen. The occupied local $\pi$ DOS of the N atom contains 1.36
electron in both the $9\times 9$ and $10\times 10$ supercells. The remaining
0.64 excess electron is distributed on the surrounding carbons, of which a total
of 0.56 sits on the three first neighbors (C1), which are therefore negatively
charged.

For both $9\times 9$ and $10\times 10$ doped supercells, there is a localized
state expelled from the $\pi$ band and located 9.35~eV below $E_F$, to be
compared with 8.65~eV in DFT calculations. This localized state weights more
than 25\% (0.57 state of the N local DOS, which can accommodate 2 electrons
including the spin degeneracy). The large value of the weight can be understood
from the arguments developed in Appendix A for a simplified tight-binding
Hamiltonian.

We conclude here that the supercell technique would require very large supercells
(up to $30\times 30$) if it were desired to describe the electronic states of an
isolated defect. In other words, the supercell technique is badly convergent
when the size of the cell increases, particularly in low dimensions. The
corresponding long range of a local perturbation produces interferences between
the images of the defect generated by the periodic boundary conditions (see also
Ref.\ [\onlinecite{Zanolli2010}]).

\section{Isolated N impurity}
\begin{figure}
\centerline{\includegraphics*[width=5cm]{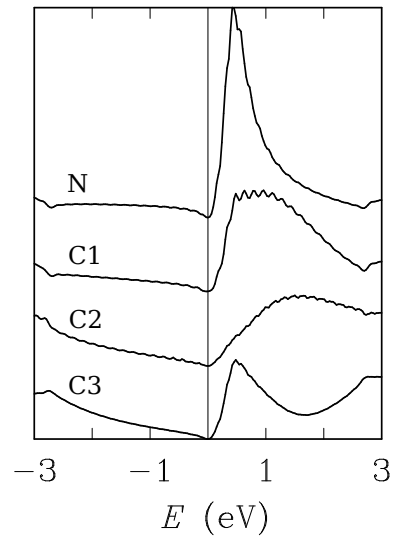}}
\caption{Tight-binding local $\pi$ densities of states on and around an isolated
  impurity in graphene. The labels correspond to the defect site (N), the first,
  second and third neighbors (C1, C2 and C3). The curves have been shifted
  vertically for clarity. The zero of energy is $\varepsilon_C$.}
\label{impur}
\end{figure}

We now turn our attention to the case of an isolated N impurity studied by
tight-binding. The local DOS on and around an isolated N impurity are presented
in Fig.~\ref{impur}. The calculations were performed on a 150x150 cell of
graphene (45,000 atoms) with periodic boundary conditions. By restricting the
number of pairs of recursion coefficients to 150, the impurities located in
adjacent cells do not feel each other, which means that the N atoms are
virtually isolated. In the N local DOS, there is a tall asymmetric peak located
0.5~eV above the Dirac energy ($\varepsilon_C$) of graphene which coincides here
with the Fermi level. The peak broadens and shifts to 1~eV on the first
neighbors and moves up further to 1.5~eV on the second neighbors. The local DOS
on the third-neighbor atoms reproduces the resonance peak of the impurity, with
a smaller amplitude. The local DOS on the fourth neighbors (not shown) bears
resemblance with that of pure graphene. It is interesting to observe in
Fig.~\ref{impur} that there remains almost nothing of the Van Hove singularities
of graphene at $\varepsilon_C \pm \gamma_0$ on the N and C1 sites.

The problem of an isolated impurity in graphene can be addressed analytically if
one simplifies the Hamiltonian to a point-like defect by ignoring the Gaussian
delocalization of the perturbation ($\sigma \to 0$ in
eq.~(\ref{gaussian})). Details are presented in Appendix A. If this simple model
captures the essential physics of the problem, it fails in producing all the
details set up by the delocalized nature of the perturbation. As shown in the
Appendix A, perturbing the first-neighbor C atoms, in addition of course to the
substitution site, leads to a better picture.

\section{Randomly distributed N atoms}
Experimentally, nitrogen doping of graphene leads to randomly distributed
substitutional sites,~\cite{Zhao2011,Joucken2012} with perhaps some preference
of the dopant atoms to sit on the same sublattice, at least
locally.~\cite{Zhao2011}  The calculations illustrated in the present Section
were performed for nitrogen distributed randomly with an atomic concentration of
0.5\%, identical to that realized with the $10\times 10$ superstructure. The
selection of the substitutional sites was constrained by the requirement that
the distance between two nitrogens remains larger than 12$\sigma$ = 1.8~nm (see
eq.~(\ref{gaussian})). The reason for that was to avoid any overlapping of the
potential wells generated by the dopants.

\begin{figure}
\centerline{\includegraphics*[height=8cm]{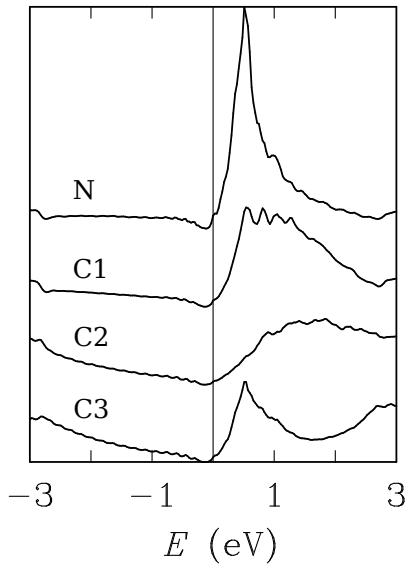}}
\caption{Configurational average of the local $\pi$ densities of states on and
  around substitutional N atoms in a graphene sheet containing 0.5~at\% of
  nitrogen atoms randomly distributed on the two sublattices. The labels
  correspond to the nitrogen dopant (N), the first, second and third neighbors
  (C1, C2 and C3). The curves have been shifted vertically for clarity. The
  zero of energy is $\varepsilon_C$.}
\label{unpave005}
\end{figure}

Fig.~\ref{unpave005} shows a configurational averaging of the local densities of
states when the N atoms have equal probabilities to sit on one or the other of
the graphene sublattices. This case will be referred to as ``unpolarized'' (or
``uncompensated'').\cite{Pereira2008} By configurational averaging, it is meant
the arithmetic average of local DOS on 50 nitrogens selected randomly --- among
the 225 dopant atoms that the 150x150 sample box contains --- and the arithmetic
average of local DOS on 150 randomly-selected C1, C2 and C3 sites. The N, C1,
C2, and C3 average local DOS for the unpolarized disordered distribution all
have a remarkable similarity with the ones obtained for an isolated
impurity (Fig.~\ref{impur}). The result are in agreement with N local DOS for
randomly distributed impurities obtained by Lherbier {\it et al.}
\cite{Lherbier2008} with a small broadening.

\begin{figure}
\centerline{\includegraphics*[height=8cm]{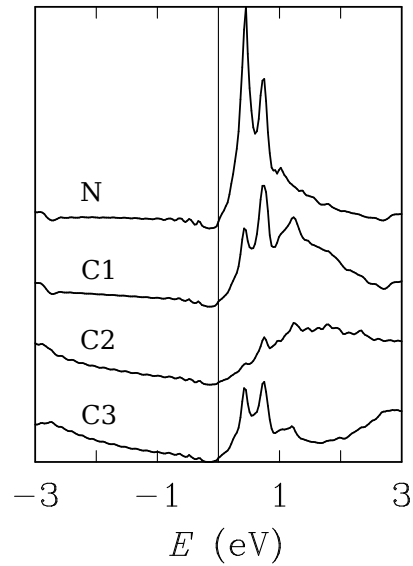}}
\caption{Configurational average of the local $\pi$ densities of states on and
  around substitutional N atoms in a graphene sheet containing 0.5~at\% of
  nitrogen atoms randomly distributed on the {\em same sublattice}. The labels
  correspond to the nitrogen dopant (N), the first, second and third neighbors
  (C1, C2 and C3). The curves have been shifted vertically for clarity.  The
  zero of energy is $\varepsilon_C$.}
\label{polave005}
\end{figure}

When the N atoms are put, still randomly, on the {\em same sublattice}
(``polarized'' case''), the multi-peak structure characteristics of the
superlattice (see Fig.~\ref{scdos}(e,f)) reappear on the average local DOS shown
in Fig.~\ref{polave005}. Interestingly, the local DOS on the second-neighbor
atoms (C2), which belongs to the same sublattice as the N atom they surround,
are virtually the same in Figs.~\ref{unpave005} and \ref{polave005}, insensitive
to the polarization of the N distribution. It is also interesting to remark
that, like with the $10\times 10$ superstructure, there is a tiny gap of states
0.1~eV below the Dirac energy. The gap looks narrower in case of the unpolarized
distribution compared to the polarized one. In the polarized case, the
appearance of a gap is natural since the average diagonal potential breaks the
symmetry between the two sublattices. Similar effects have been observed in the
case of vacancies.\cite{Pereira2008}

\begin{figure}
\centerline{\includegraphics*[width=7cm,angle=90]{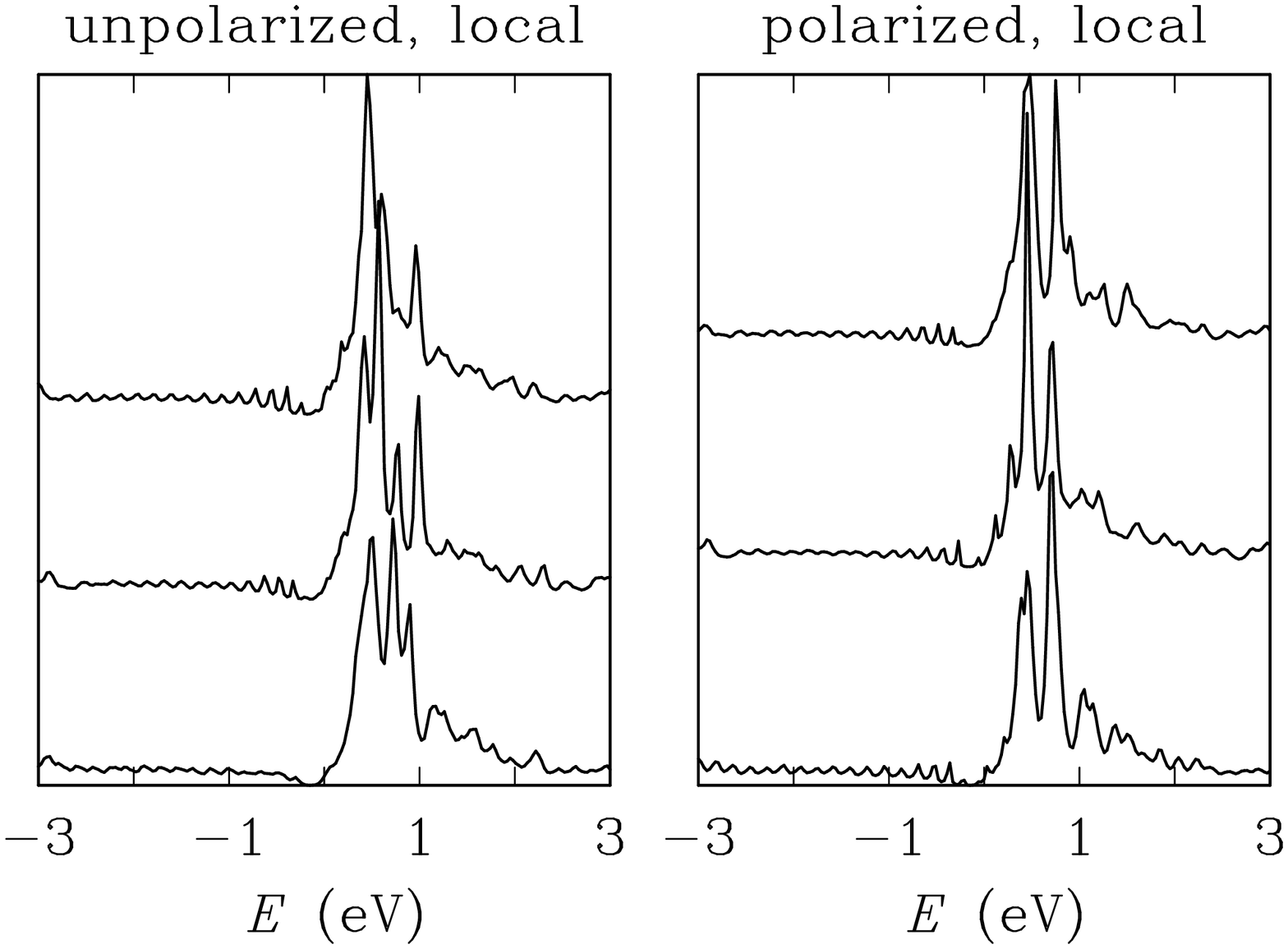}}
\caption{Local DOS on three N atoms randomly selected in graphene doped at
  concentration 0.5 at\%. The unpolarized case (left) refers to dopants located
  on the two sublattices, whereas the dopants all sit on the same sublattice in
  the polarized case (right).  The zero of energy is $\varepsilon_C$.}
\label{bothlocal}
\end{figure}

Averaging the local densities of states makes visible the distinction between
unpolarized and polarized distributions of N. However, for a given configuration
of the dopant atoms, the local DOS varies from site to site. How much that
variation can be is illustrated in Fig.~\ref{bothlocal}, which compares local
DOS calculated on three different N atoms for both the unpolarized and the
polarized distributions. The variations from site to site are so important that
the local DOS is not a reliable indicator of the global distribution of the
nitrogen atoms among the two sublattices. If it is true that the N local DOS
displayed in Fig.~\ref{bothlocal} are less peaked that the N local DOS of the
$10\times 10$ superstructure (Fig~\ref{scdos}(e,f)), identifying which is which
would be difficult on the experimental side if one had only STS spectra to deal
with.

 Finally, it may be important to remark that the minimum of the $\pi$ density of
 states around the N dopants does not coincide with the atomic level
 $\epsilon_C$ of carbon in graphene (the Dirac point energy; see also Refs.
 [\onlinecite{Pereira2008}, \onlinecite{Skrypnyk2009}, \onlinecite{Wu2008}]).
 Depending on the concentration, the DOS minimum shifts slightly below
 $\epsilon_C$ (see e.g. Figs.~\ref{unpave005} and ~\ref{polave005}).

\section{STM images}
A reliable information on the electronic structure of doped graphene can be
obtained by STM imaging. A tip polarized negatively compared to the graphene
layer probes the unoccupied states where a N substitutional impurity and the
adjacent carbon atoms have peaks in their local DOS. The increase of electronic
density of states, compared to graphene, in this energy region gives rise to a
bright triangular spot in the STM
image,~\cite{Zheng2010,Zhao2011,Joucken2012,Brito2012} with two possible
orientations with respect to the honeycomb lattice depending on the sublattice
on which the N dopant sits. These two orientations are rotated by 180$^{\circ}$
from each other, as observed experimentally.\cite{Zhao2011,Joucken2012}

\begin{figure}
\centerline{\includegraphics*[width=7cm]{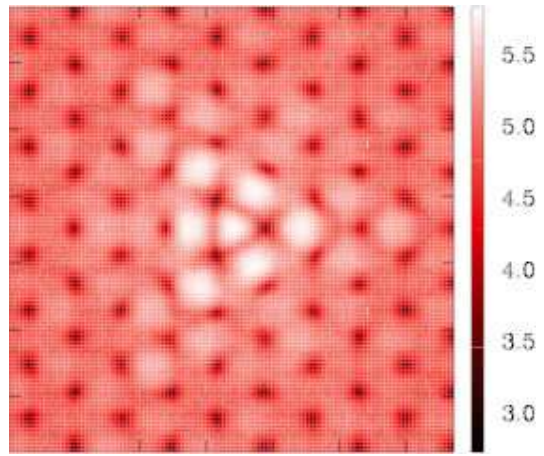}}
\caption{Tight-binding STM constant-current image of graphene with an isolated N
  impurity located near the center, computed for a negatively-polarized tip
  ($V_{\text{tip}}$ = -0.5~V). The vertical scale is the tip height (\AA) above
  sample. The image size is 2~nm along both sides.}
\label{stmimpurneg}
\end{figure}

Fig.~\ref{stmimpurneg} is a tight-binding STM image~\cite{Meunier1998} computed
for graphene with a single N impurity. The prominent triangular arrow-head at
the center of the image is located on the N, C1 and C3 sites. The
second-neighbor carbons (C2) atoms do not participate much to the STM signature
of the dopant. This is so because the local DOS in the energy window probed by
the STM current (0 to 0.5~eV) is smaller on the C2 atoms than on the N, C1 and
C3 atoms (see Fig.~\ref{impur}).

\begin{figure}
\centerline{\includegraphics*[width=7cm]{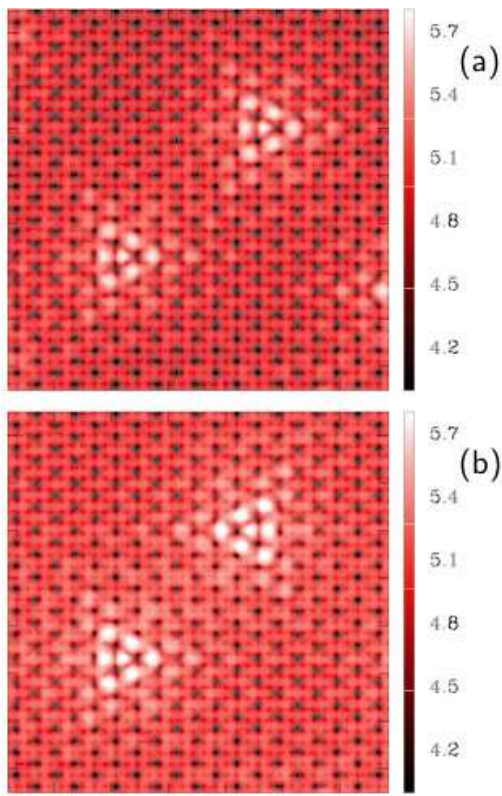}}
\caption{Constant-current STM images computed for two configurations of graphene
  doped with 0.5~at\% N. The distribution of the N atoms is random. The two
  nitrogen atoms visible in image (a) lie on the same sublattice, which is not
  the case in image (b). In both cases, the STM tip is polarized at -0.5~V
  with respect to the sample. Each image is a square of 4~nm edge.}
\label{stm005}
\end{figure}

Fig.~\ref{stm005} shows two STM images computed for graphene doped with N at
0.5~at\% concentration, the dopants being randomly distributed on the two
sublattices. In the configuration shown in image (a), there are two impurities
located 1.9~nm apart sitting on the same sublattice. The two nitrogens captured
in images (b) do not sit on the same sublattice, which explains why the related
triangular STM patterns are oriented differently. The scale used for the STM
signal (tip height at constant current) is identical for both images. For
improving the contrast, the scale has been saturated below 4.0~\AA\ and above
5.8~\AA. Two dopants interfere more when they are located on the two
sublattices than when they sit on the same sublattice.

\begin{figure}
\centerline{\includegraphics*[width=7cm]{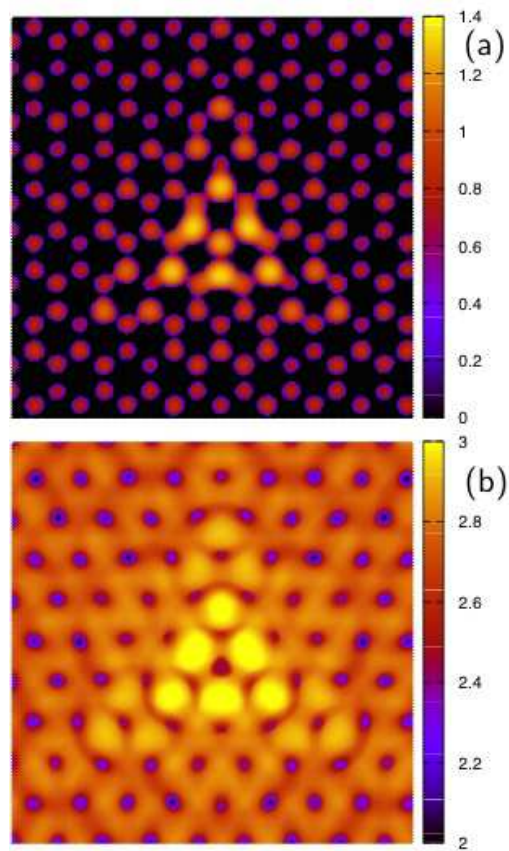}}
\caption{DFT calculation of constant-current STM images of $10\times 10$ N-doped
  graphene for two levels of the tunnel current: (a) high and (b) low (see
  text). The tip bias polarization is -0.5~V, the vertical scale is the tip
  height (\AA) above sample. Each image represents a square of 4~nm edge
  centered on a N dopant.}
\label{stmdft}
\end{figure}

In tight-binding STM theory,~\cite{Meunier1998} there is an empirical parameter
coupling the tip apex to the sample atoms, which was chosen independent of the
chemical nature of the probed atoms. In other words, the STM calculations
differentiate a N atom from the C atoms only through intrinsic effects that the
former has on the electronic structure of the doped graphene. In DFT
calculations, the standard way to generate an STM image is via Tersoff-Hamann's
theory.~\cite{Tersoff1983} In addition to local variations of the DOS, the
tunnel current depends on the spatial extension of the $\pi$ orbitals in the
direction perpendicular to the atomic plane. For doped graphene, the $2p_z$
orbital of N decreases more rapidly than the C $2p_z$ orbital
does,~\cite{Zheng2010} which means that the contrast of the computed image
depends on the tip--sample distance. This dependence is illustrated in
Figs.~\ref{stmdft}(a) and (b) that represent DFT constant-current STM images
computed for a small current (large tip-sample distance) and for a large current
(small tip-sample distance), respectively. The local DOS is larger on top of the
N atom than on the top of all the other atoms, but it decreases much more
rapidly in the normal direction. This explains the small brightness of the N
site for the large tip--sample distance. Experimentally, variations of the STM
contrast have also been observed as a function of the current/voltage
conditions,~\cite{Joucken2012} where image of the dopant atom protrudes in some
occasions and does not in other occasions. It is tempting to attribute this
observation to variations of the tip--sample distances depending on the current
setpoint, as in Fig.~\ref{stmdft}.  However, the experimental distance is much
larger than the one that can be achieved numerically in DFT, due to the fast
decay of the localized basis set when moving away from the sample surface.

\section{Conclusions}

Tight-binding and DFT electronic calculations reveal that the perturbation
induced by N dopants in graphene is long ranged. This is not a surprise when one
remembers that the Green function of the free-electron problem in two dimensions
is the Bessel-Hankel function of zeroth order
$(i/4)H_0^{(1)}(k|\vec{r}-\vec{r}'|)$, which slowly decreases like $1/\sqrt{kd}$
with the distance $d = |\vec{r}-\vec{r}'|$, where $k = \sqrt{\hbar^2/2mE}$.  For
the $\pi$-electron tight-binding Hamiltonian on the honeycomb lattice, the Green
function elements $G_{nm}^0$ can also be approximated, at low energy, by Hankel
functions of $\delta k|\vec{r}_m - \vec{r}_n|$, now of order zero or one,
depending on whether they refer to sites $m$ and $n$ located on the same
sublattice or not.\cite{Wang2006} In this regime, the energy $E$ is linearly
related to $\delta k$ by the dispersion relation $E=\hbar v_F \delta k$, where
$v_F$ is the Fermi velocity, and the Hankel functions are also multiplied by the
energy and by periodic functions describing $\sqrt{3}\times\sqrt{3}$ modulations
related to the Dirac wave vectors $K$ and $K'$. For large separation distances,
the Green function elements decay slowly, like $1/\sqrt{E\;|\vec{r}_m
  -\vec{r}_n|/\hbar v_F}$, except for very small excitation energies where the
Hankel functions diverge. A more detailed discussion will be given
elsewhere. \cite{FD2012}

The scattering formalism developed in Appendix A for the case of a point-like
defect emphasizes the central role of the Green function in the understanding of
the perturbation induced by the defect. Hence the long-range interaction between
dopants in graphene. It is important to realize that this long-range interaction
is not the consequence of using an extended perturbation of the on-site energies
around the defect (eq.~(\ref{gaussian})). Things actually go the other way
round: a defect perturbs the crystal potential far away, because of the slow
decay of the scattering it produces. As a consequence, the parameters of the
tight-binding Hamiltonian that mimic {\it ab-initio} calculations are affected
in a sizable neighborhood of the defect.~\cite{Adessi2006}

This long-range interaction has several effects. First, it produces
interferences between the duplicates of the defect generated by periodic
boundary conditions when a supercell approach is used. The latter must therefore
be used with caution. Second, substitutional impurities dispersed in graphene
cannot be treated as independent defects as soon as the distance between two of
them becomes too small, about 2~nm in the case of nitrogen. Increasing the
strength of the local potential also increases the range of the perturbation it
produces. This unusual behavior is actually related to the properties of the
Green functions at low energy (see Appendix A and
Refs. [\onlinecite{Skrypnyk2009}, \onlinecite{Skrypnyk2006},
  \onlinecite{Skrypnyk2011}]). As mentioned in Appendix A, the deepest defect is
the vacancy. Nitrogen, with its 10~eV effective perturbation parameter, is not a
shallow defect. Boron would not be a shallow impurity either. B substitution can
be described by a potential hump, instead of a well, with a similar $|U|$ as
nitrogen and a similar long-range perturbation.~\cite{Latil2004,Khalfoun2010} In
the case of more complex defects, such as N plus vacancy,~\cite{Hou2012} P plus
N,~\cite{Cruz2009} or O$_3$,~\cite{Leconte2010} it would be more difficult to
define local parameters and to assess the importance of long-range interactions
along the same way as in this paper.

As demonstrated experimentally, the partition of the honeycomb network in two
sublattices has subtle effects on STM image of graphene with substitutional
impurities, as exemplified by Fig.~\ref{stm005}. The STM image of a dopant in
graphene may be influenced by the proximity of another dopant. This is a direct
consequence of the defect interactions mediated by graphene. What happens when
two impurities come very close to each other remains to be clarified.  A new
parametrization of the tight-binding Hamiltonian would be required if one had to
obtain, from calculations, the STM topography of neighboring N dopants.

{\bf Acknowledgments} This research has used resources of the Interuniversity
Scientific Computing Facility located at the University of Namur, Belgium, which
is supported by the F.R.S.-FNRS under convention No. 2.4617.07 and the FUNDP.

\section*{Appendix A. The impurity problem in graphene}
In this Appendix, a simple but illustrative perturbation model is developed. It
is assumed that the on-site energies all take the unperturbed bulk value
$\varepsilon_C = 0$, except on the nitrogen atom where $\varepsilon_N = U$. The
tight-binding Hamiltonian can be set in the form:
\begin{equation}
H = H^0 + V = -\sum_{n,m} |n\rangle t _{nm} \langle m|  + |0\rangle U \langle 0|\; . \label{H0V}
\end{equation}
The states $|n\rangle $ denote $\pi$ orbitals centered on sites $n$:
$\varphi_{\pi}(r-n) = \langle r|n\rangle$. $H^0$ is the usual tight-binding
Hamiltonian where only hopping integrals $t_{nm}\equiv\gamma_0$ between first
neighbors $n$ and $m$ are kept, and $V$ is the localized potential of the N atom
at site $0$. The local density of states $n(r,E) $ is given by:
\begin{equation}
n(r,E)=\sum_{n,m} n_{nm}(E)\varphi_\pi(r-n)\varphi_\pi(r-m)\;,
\end{equation}
where $n_{nm}(E)$ is obtained from the Green function or resolvent
$G(z)=(z-H)^{-1}$:
\begin{equation}
n_{nm}(E)= \frac{-1}{\pi} \; \lim_{\varepsilon \to 0} \text{ Im } \langle n|G(E+i\varepsilon)|m\rangle \; .
\end{equation}
$G(z)$ can then be calculated in terms of the unperturbed Green function
$G^0(z)=(z-H^0)^{-1}$; this is the so-called Koster-Slater-Lifshitz problem. If the
matrix elements $ \langle n|G(E+i\varepsilon)|m\rangle$ are written $G_{nm}(z)$,
we have:
\begin{equation}
G_{nm} = G^0_{nm} + G^0_{n0}\tau G^0_{0m} \quad; \quad \tau=\frac{U}{1-UG^0_{00}}\; . 
\label{eqgreen}
\end{equation}
In particular, on the impurity site, we have
$G_{00}=G^0_{00}/(1-UG^0_{00})$,\cite{KosterSlater1954,Lifshitz1965} and the
local density of states on the N site, $n_N(E)\equiv n_{00}(E)$ is given by:
\begin{equation}
n_N(E)=\frac{ n^0(E)}{(1-UF^0(E))^2 +\pi^2U^2n^0(E)^2}\;,
\label{nN(E)}
\end{equation}
where $F^0(E)$ is the real part of the Green function
$G^0_{00}(E+i\varepsilon)$, {\it i.e.} the Hilbert transform of the unperturbed
density of states of graphene $n^0(E)$.

\begin{figure}
\centerline{\includegraphics*[width=7cm]{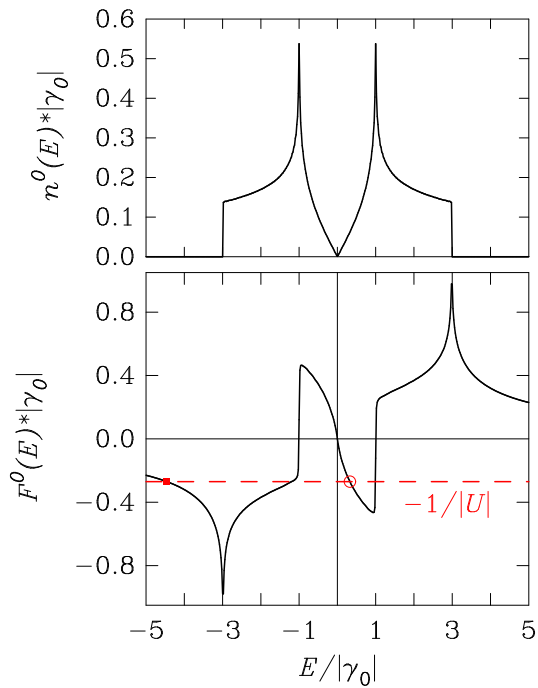}}
\caption{Real ($F^0(E)$) and imaginary ($n^0(E)$) parts of the Green function of
  the graphene $\pi$ tight-binding Hamiltonian along the real axis, in units of
  $|\gamma_0|$. The intersection of $F^0(E)$ with the reciprocal of the
  perturbed level $1/U$ (dashed line) in the vicinity of the Dirac point gives
  the energy $E_r$ at which the local DOS of the impurity has a resonance (open
  circle symbol). There is another intersection (filled symbol) giving rise to a
  localized state below the $\pi$-electron density of states. }
\label{green}
\end{figure}

\begin{figure}
\centerline{\includegraphics*[width=8cm]{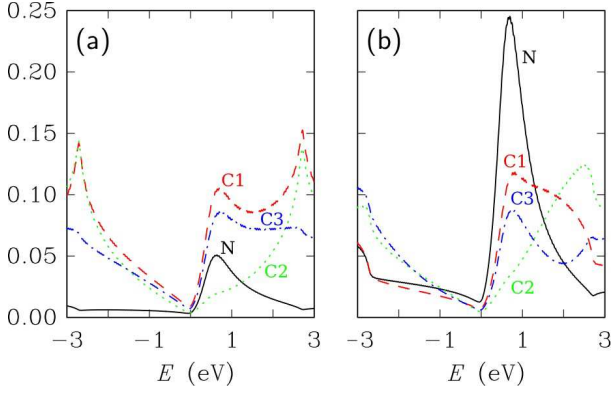}}
\caption{Local densities of states deduced from the simple tight-binding model
  for an isolated nitrogen impurity in graphene. (a) On-site perturbation energy
  $U$ = -10~eV on the N site only; (b) on-site perturbation on the N (-4~eV) and
  the three C1 sites (-2.57~eV).}
\label{simpmod}
\end{figure}

Close to the Dirac point, $n^0(E) \sim |E|$ and $F^0(E) \sim E \ln |E|$, and
provided $U$ is large enough, $n_N(E)$ shows a resonant behavior around the
energy $E_r$ such that $(1-UF^0(E_r))=0$ (see the open-circle symbol in
Fig.~\ref{green}). This has been discussed in many places.~\cite{Skrypnyk2006,
  Wang2006, Wehling2007, Pereira2008, Wu2008, Toyoda2010, Chang2011} It turns
out that a value for $U$ of the order of $-10$~eV reproduces a resonance at the
position obtained with the previous model (compare Fig.~\ref{simpmod}(a) with
Fig.~\ref{impur}) instead of the value $U = -4$ eV in the previous
approach. However, the shapes of the local DOS differ to some extent between the
local-perturbation model developed in this Appendix and the more delocalized
potential well considered throughout this paper. The resonance here
(Fig.~\ref{simpmod}(a)) is weaker on the N site than on the sites C1 and C3
located on the other sublattice, which can be explained from the symmetry
properties of the unperturbed Green functions: $G_{nm}(z)$ is and odd or even
function of $z$ depending on whether the sites $n$ and $m$ belonging to the same
sublattice of the graphene structure or not. As a consequence, for example,
$G^0_{00}(z=0)=0$ whereas the real part of $G^0_{0n}(z=0)$ does not vanish when
$n$ is a first neighbor, which has a big impact on the local densities of states
(see eq.\ (\ref{eqgreen})).~\cite{FD2012} Also the Van Hove singularities at
2.7~eV are still present here.

In addition to the {\em resonant} state located above the Dirac point,
Fig.~\ref{green} reveals the existence of a {\em localized} state lying below
the $\pi$-electron DOS, where the on-site perturbation level $1/U$ intersects
the curve $F^0(E)$ at $-4.5|\gamma_0|$ (solid square symbol). The weight
(residue) of this localized state is inversely proportional to the slope of
$F^0(E)$ at the intersection. This is the localized state found at 9.35~eV below
$E_F$ in tight-binding and 8.65~eV below $E_F$ in DFT calculations for the doped
superstructures (Section~\ref{section_periodic}).

This one-parameter model can be generalized to other type of local defects or
chemical doping. B doping, with a positive $U$, will yield results symmetrical
to those obtained for N: it can indeed be anticipated from Fig.~\ref{green} that
the resonance state will now appear near the top of the occupied states, as confirmed by DFT calculations.~\cite{Zheng2010,Latil2004,Khalfoun2010} The
limit $U \to \infty$ corresponds to the introduction of a vacancy instead of a
substitutional impurity. Because of the vanishing of the density of states and
of the logarithmic divergence of $F^0(E)$, this limit is singular and must been
studied with special
care.\cite{Pereira2008,Wu2008,Toyoda2010,Chang2011,Yuan2010} The resonant state
becomes very narrow, but does not become a genuine bound state. The total
density of states varies as $1/([\ln|E|]^2|E|)$ and the perturbation of the
electronic density is concentrated on the first neighbors and on the sites
belonging to the sublattice different from that of the vacancy.\cite{FD2012}

The simple model developed here above allows one to address the question of the
role of N-C hopping interactions on the local DOS on the impurity. When the N-C
hopping takes a value $\gamma$ different from the one $\gamma_0$ between the C
atoms, the Green function element on the perturbed site can be set in the form
\begin{equation}
G_{00}(z) = \frac{1}{z-U-(\gamma^2/\gamma_0^2)\Sigma^0(z)}
\label{sigma}
\end{equation} 
where $\Sigma^0(z)$ is the self energy of the unperturbed graphene. In the
notations of the above formalism, the self energy can be identified to
$\Sigma^0(z) = z - 1/G^0_{00}(z)$. Inserting this expression in the right-hand
side of eq.~(\ref{sigma}) leads to a generalization of eq.~(\ref{nN(E)}) valid
for $\gamma \neq \gamma_0$. In particular, the resonance condition in the
vicinity of $E$ = 0 becomes
\begin{equation}
\frac{\gamma^2}{\gamma_0^2 U + (\gamma^2 - \gamma_0^2)E_r} = F^0(E_r)\;.
\end{equation} 
By comparison with the situation depicted in Fig.~\ref{green}, the intersection
of $F^0(E)$ needs now to be search for with a curve whose ordinate and slope at
the origin are $-\gamma^2 /(\gamma_0^2|U|)$ and $\gamma^2 (\gamma_0^2 -
\gamma^2) /(\gamma_0^2 U)^2$, respectively. If $|\gamma| < |\gamma_0|$, which
should be the case here since the N atom is smaller than the C one, the ordinate
at the origin of the curve moves up and its slopes becomes positive. These two
effects pull the resonant energy $E_r$ closer to the origin compared with the
simplest situation where $|\gamma| = |\gamma_0|$. However, if the hopping
perturbation is small, renormalizing the on-site level $U$ to a larger absolute
value $U_{\text{eff}}$ would produce the same effect. This conclusion validates
the approach used so far to not modifying the N-C hopping.

We have finally considered an intermediate model between the Gaussian
distribution of the on-site energies and the one-parameter model just
described. We took $U_0$ on the N site and a second perturbation potential $U_1$
on the first neighbors, with values fixed by eq.\ (\ref{gaussian}), $U_0=-4$ eV;
$U_1=-2.57$ eV. The agreement with the full Gaussian model is much better than
with the one parameter model (Fig.~\ref{simpmod}(b)). It is fairly remarkable to
realize that the intensity of the resonance state on the N site increases when
the potential is delocalized on the first neighbors.

\section*{Appendix B. Band gap opening in doped superstructures}

The following perturbation theory is based on the same model Hamiltonian as in
eq.~(\ref{H0V}), except that the potential $V$ refers now to the perturbation
brought about by the periodic array of nitrogen atoms in substitution for C in a
$p\times p$ supercell of graphene. These atoms occupy one among the $2N = 2p^2$
sublattices of the supercell, here denoted sublattice $1$:
\begin{equation}
H = H^0 + V = -\sum_{n,m} |n\rangle t _{nm} \langle m|  + \sum_{n \in 1} |n\rangle U \langle n|\; .
\end{equation}
When $V=0$, there are four states with zero energy (Dirac energy of the
unperturbed graphene): $|K^i\rangle$ and $|K'^i\rangle$, where $i=A,B$ denote
the sublattices of the graphene structure and $|K^i\rangle$, $|K'^i\rangle$ are
the corresponding Bloch functions:
\begin{equation}
|K^{A(B)}\rangle = \frac{1}{\sqrt{N}} \sum_{n\in A(B)} e^{i K.n} |n\rangle \; .
\end{equation}
where $K.n$ in the exponential represents the dot product of the wave vector of
the $K$ point of graphene and the position vector of the site $n$ in real space.

It will be assumed that the sublattice $1$ is an $A$ sublattice. Then, the
states $|K^{B}\rangle$ and $|K'^{B}\rangle$, having no amplitude on sublattice
$1$, are eigenstates of zero energy for any value of the on-site perturbation
$U$. Very close to zero energy, furthermore, only the ($(|K^{A}\rangle,
|K'^{A}\rangle)$) subspace needs to be considered. In particular,
\begin{eqnarray}
\langle K^A|V|K^A \rangle &=& \langle K'^A|V|K'^A \rangle = \frac{1}{N} \sum_{n\in 1} 1 = U/p^2 \\
\langle K^A|V|K'^A \rangle &=& \frac{1}{N} \sum_{n\in 1} \exp[i(K'-K).n] = U/p^2 \, \delta_{K'-K,G} \; ,
\end{eqnarray}
where the Kronecker symbol imposes that $K'-K$ be a vector of the reciprocal
lattice of the supercell, \textit{i.e.} it is one when $p=3q$ is a multiple
integer of 3 and is $0$ if $p$ is not divisible by 3.

As a consequence, in the case $p=3q$, the four states with zero energy are
mapped onto the zone center. Two of them, with non-zero amplitudes on the $B$
sites, remain degenerate with zero energy; the other two states have energy
$E=0$ and $E=-2|U|/p^2$ to lowest order in perturbation. There are then two plus
one states at zero energy, but this is an accidental degeneracy due to the
simplified tight-binding model. When $p$ is not divisible by 3, the degeneracy at the $K$ and $K'$ points of the folded zone is lifted: at each point, one state remains at zero energy
($|K^{B}\rangle$ or $|K'^{B}\rangle$, respectively) whereas the other state
moves down at energy $-|U|/p^2$ to lowest order in perturbation.

All these features are clearly visible in the band structures of the $9\times 9$
and $10\times 10$ superlattices shown in Fig.~\ref{bstrdft}. Since a realistic
value for $|U|$ is about 10~eV, the gap for the $10\times 10$ case should be of
the order of $10/p^2 = 0.1$~eV, not far from the actual value (see Section
III). For the $9\times 9$ superstructure, the expected crossing between the two
linear $\pi$ and $\pi^*$ bands is observed at the $\Gamma$ point close to two
parabolic branches separated from each other by a gap about twice this value.

\end{document}